\documentclass[ aps,prl,twocolumn,superscriptaddress,floatfix,nofootinbib,showpacs,amsmath,amssymb,altaffilletter]{revtex4-1}

\usepackage[left]{lineno}

\usepackage[colorlinks=true,pdfstartview=FitV,linkcolor=blue,citecolor=blue, 
urlcolor=blue]{hyperref}
\usepackage{graphicx}

\newcommand{\APC}{APC, Universit\'e Paris Diderot, CNRS/IN2P3, CEA/Irfu, Obs de Paris, USPC, Paris 75205, France}
\newcommand{\AQLNGS}{INFN Laboratori Nazionali del Gran Sasso, Assergi (AQ) 67100, Italy}
\newcommand{\AQGSSI}{Gran Sasso Science Institute, L'Aquila 67100, Italy}

\newcommand{\Augustana}{Physics Department, Augustana University, Sioux Falls, SD 57197, USA}
\newcommand{\Belgorod}{Radiation Physics Laboratory, Belgorod National Research University, Belgorod 308007, Russia}
\newcommand{\BHSU}{School of Natural Sciences, Black Hills State University, Spearfish, SD 57799, USA}

\newcommand{\CAUniPHY}{Physics Department, Universit\`a degli Studi di Cagliari, Cagliari 09042, Italy}
\newcommand{\CAINFN}{INFN Cagliari, Cagliari 09042, Italy}

\newcommand{\Campinas}{Physics Institute, Universidade Estadual de Campinas, Campinas 13083, Brazil}
\newcommand{\CentroFermi}{Museo della fisica e Centro studi e Ricerche Enrico Fermi, Roma 00184, Italy}

\newcommand{\CTLNS}{INFN Laboratori Nazionali del Sud, Catania 95123, Italy}
\newcommand{\ENUniCEE}{Engineering and Architecture Faculty, Universit\`a di Enna Kore, Enna 94100, Italy}

\newcommand{\FNAL}{Fermi National Accelerator Laboratory, Batavia, IL 60510, USA}

\newcommand{\GEUni}{Physics Department, Universit\`a degli Studi di Genova, Genova 16146, Italy}
\newcommand{\GEINFN}{INFN Genova, Genova 16146, Italy}

\newcommand{\Hawaii}{Department of Physics and Astronomy, University of Hawai'i, Honolulu, HI 96822, USA}
\newcommand{\Houston}{Department of Physics, University of Houston, Houston, TX 77204, USA}
\newcommand{\IHEP}{Institute of High Energy Physics, Beijing 100049, China}

\newcommand{\JINR}{Joint Institute for Nuclear Research, Dubna 141980, Russia}
\newcommand{\Krakow}{M. Smoluchowski Institute of Physics, Jagiellonian University, 30-348 Krakow, Poland}
\newcommand{\Kurchatov}{National Research Centre Kurchatov Institute, Moscow 123182, Russia}

\newcommand{\LPNHE}{LPNHE, CNRS/IN2P3, Sorbonne Universit\'e, Universit\'e Paris Diderot, Paris 75252, France}

\newcommand{\MEPhI}{National Research Nuclear University MEPhI, Moscow 115409, Russia}

\newcommand{\MIINFN}{INFN Milano, Milano 20133, Italy}

\newcommand{\MIUni}{Physics Department, Universit\`a degli Studi di Milano, Milano 20133, Italy}
\newcommand{\MSU}{Skobeltsyn Institute of Nuclear Physics, Lomonosov Moscow State University, Moscow 119234, Russia}
\newcommand{\NAINFN}{INFN Napoli, Napoli 80126, Italy}
\newcommand{\NAUniPHY}{Physics Department, Universit\`a degli Studi ``Federico II'' di Napoli, Napoli 80126, Italy}

\newcommand{\Petersburg}{Saint Petersburg Nuclear Physics Institute, Gatchina 188350, Russia}
\newcommand{\PGUniCBB}{Chemistry, Biology and Biotechnology Department, Universit\`a degli Studi di Perugia, Perugia 06123, Italy}
\newcommand{\PGINFN}{INFN Perugia, Perugia 06123, Italy}
\newcommand{\PIINFN}{INFN Pisa, Pisa 56127, Italy}
\newcommand{\PIUniPHY}{Physics Department, Universit\`a degli Studi di Pisa, Pisa 56127, Italy}
\newcommand{\PNNL}{Pacific Northwest National Laboratory, Richland, WA 99352, USA}
\newcommand{\Princeton}{Physics Department, Princeton University, Princeton, NJ 08544, USA}

\newcommand{\RMTreINFN}{INFN Roma Tre, Roma 00146, Italy}
\newcommand{\RMTreUni}{Mathematics and Physics Department, Universit\`a degli Studi Roma Tre, Roma 00146, Italy}
\newcommand{\RMUnoINFN}{INFN Sezione di Roma, Roma 00185, Italy}
\newcommand{\RMUnoUni}{Physics Department, Sapienza Universit\`a di Roma, Roma 00185, Italy}

\newcommand{\SSUniCHP}{Chemistry and Pharmacy Department, Universit\`a degli Studi di Sassari, Sassari 07100, Italy}

\newcommand{\Temple}{Physics Department, Temple University, Philadelphia, PA 19122, USA}

\newcommand{\UCDavis}{Department of Physics, University of California, Davis, CA 95616, USA}
\newcommand{\UCLA}{Physics and Astronomy Department, University of California, Los Angeles, CA 90095, USA}
\newcommand{\UMass}{Amherst Center for Fundamental Interactions and Physics Department, University of Massachusetts, Amherst, MA 01003, USA}

\newcommand{\USP}{Instituto de F\'isica, Universidade de S\~ao Paulo, S\~ao Paulo 05508-090, Brazil}
\newcommand{\VTech}{Virginia Tech, Blacksburg, VA 24061, USA}

\begin{document}
\title{Effective field theory interactions for liquid argon target in DarkSide-50 
experiment}

\author{P.~Agnes}\affiliation{\Houston}
\author{I.F.M.~Albuquerque}\affiliation{\USP}
\author{T.~Alexander}\affiliation{\PNNL}
\author{A.K.~Alton}\affiliation{\Augustana}
\author{M.~Ave}\affiliation{\USP}
\author{H.O.~Back}\affiliation{\PNNL}
\author{G.~Batignani}\affiliation{\PIINFN}\affiliation{\PIUniPHY}
\author{K.~Biery}\affiliation{\FNAL}
\author{V.~Bocci}\affiliation{\RMUnoINFN}
\author{G.~Bonfini}\affiliation{\AQLNGS}
% ...
\collaboration{The DarkSide-50 Collaboration}\noaffiliation

\begin{abstract}
We reanalize data collected with the DarkSide-50 experiment and recently used to set limits on the spin-independent interaction rate of weakly interacting massive particles (WIMPs) on argon nuclei with an effective field theory framework. The dataset corresponds to a total (16660 $\pm$ 270) kg d exposure using a target of low-radioactivity 
argon extracted from underground sources. We obtain upper limits on the effective couplings of the 12 leading operators in the nonrelativistic systematic expansion. 
For each effective coupling we set constraints on WIMP-nucleon cross sections, setting upper limits between $2.4 \times 10^{-45} \, \mathrm{cm}^2$ and 
$2.3 \times 10^{-42} \, \mathrm{cm}^2$ (8.9 $\times 10^{-45} \, \mathrm{cm}^2$ and 6.0 $\times 10^{-42} \, \mathrm{cm}^2$) for WIMPs of mass of 100 $\mathrm{GeV/c^2}$ (1000 $\mathrm{GeV/c^2}$) at 90\% confidence level.

\vspace{1cm}
\end{abstract}

\pacs{To be defined}
\maketitle

\section*{Introduction}

Astrophysical and cosmological observations show that most of the 
matter in the Universe is  dark and nonbaryonic, whose intrinsic nature is 
still unknown~\cite{bib:dm1,bib:dm2,bib:dm3}. Compelling theoretical 
models assume that dark matter
consists of weakly interacting massive particles (WIMPs), a simple 
hypothesis able to explain the most crucial phenomenology~\cite{bib:dm4} with relative ease, like rotation curves of
spiral galaxies, the observations of anisotropies of
the cosmic microwave background, gravitational lensing at galactic scale, and the big-bang nucleosynthesis. Present theoretical research 
describes the interaction between WIMPs and target nuclei 
in terms of effective field theory (EFT) operators
\cite{bib:eft1,Fitzpatrick:2012ix,Anand:2013yka}. The lowest-order
term in a systematic nonrelativistic expansion is an
interaction that does not depend on
the relative velocity $v$ of the incoming particle or on the momentum
transfer $\vec{q}$, which can be parametrized by
spin-independent (SI) and spin-dependent cross sections. The SI cross
section is the only one relevant for spin-zero nuclei and, if WIMPs interact
coherently with all nucleons, it is 
enhanced by a factor equal to the mass number $A$
relative to incoherent cross sections like the spin-dependent cross section.

The standard SI WIMP-nucleus interaction in the galactic standard halo
scenario~\cite{bib:halo1,bib:halo2,bib:halo3} is the benchmark that is used to 
compare different experiments. The physical interpretation of the observed results 
changes under different hypotheses for the interaction.
Such a consideration is important given the present unclear experimental  
landscape. On the one hand,
 DAMA~\cite{bib:dama,Bernabei:2018yyw}
 recorded  a signal  that is interpreted as collisions of WIMPs with mass of 
a few tens of $\mathrm{GeV/c^2}$ and the CDMS II-Si~\cite{Agnese:2013rvf}
result appears to be better fitted by a model with WIMPs than by one
with only reasonable backgrounds.  On the other hand, the lack of signals
in other experiments, such as Xenon100~\cite{bib:xenon}, LUX~\cite{Akerib:2017kat}, PANDAX-II~\cite{Cui:2017nnn}, and XENON1T~\cite{Aprile:2018dbl}
seems to contradict the existence of WIMPs of this mass, if the SI interaction is coherent and
independent of the nucleus
\cite{bib:halo4}. 
WIMP-nucleus interactions that differ from the
lowest-order SI one could alleviate the tension between  experiments that use different target
nuclei. In fact, cross sections from other operators can depend on characteristics 
of the target nuclei besides the mass number $A$. In particular, they can uniquely
depend on the WIMP mass and velocity yielding interaction rates that span many
orders of magnitude~\cite{bib:nsi1,bib:nsi2,bib:nsi3,bib:op16,Aprile:2017aas,Xia:2018qgs}.

In this work, we briefly review the main ideas underlying a 
general classification of operators and form factors that can appear in WIMP-nucleus interactions.
We then focus on an argon target and, specifically, to the DarkSide-50 dataset~\cite{Agnes:2018fwg}.

\section{Effective field theory expansion for liquid argon nuclei} \label{sec:nsi}
Following the model independent approach to WIMP-nucleus scattering that uses
a Galilean-invariant EFT and the notation of Ref.~\cite{Anand:2013yka},
the interaction between two particles with nonzero 
masses can be reduced to a linear 
combination of 15 operators, if we assume, in analogy with the standard  analysis for the SI interaction, that coupling
coefficients $c_i$ are 
equal for protons and neutrons (isospin independent interaction):
\begin{equation}
\label{eq:OperatorSum}
   \mathcal{O}_{\rm int} \equiv \sum_{i=1}^{15} c_i \mathcal{O}_i \quad .
\end{equation}
This assumption makes it possible to compare limits from experiments that use different target nuclei. Providing limits on specific dynamical WIMP interaction models or
combining future positive WIMP signals from different target nuclei to gain information on 
the isospin content of the interaction requires twice as many operators and 
corresponding couplings.

Seven operators contribute to the nuclear matrix elements of the
interaction of a WIMP with the spin-zero nucleus of $^{40}$Ar:
\begin{eqnarray} \label{eq:ops}
\mathcal{O}_1 & = & 1_{\chi} 1_N  \nonumber \\	
\mathcal{O}_3 & = & i \vec{S}_N \cdot \left(  \frac{\vec{q}}{m_N} \times 
\vec{v}^{\perp} \right) \nonumber \\
\mathcal{O}_5 & = & i \vec{S}_\chi \cdot \left(  \frac{\vec{q}}{m_N} \times 
\vec{v}^{\perp} \right) \nonumber \\	
\mathcal{O}_8 & = & \vec{S}_\chi \cdot \vec{v}^{\perp} \nonumber \\	
\mathcal{O}_{11} & = & i \vec{S}_\chi  \cdot \frac{\vec{q}}{m_N}  \nonumber \\	
\mathcal{O}_{12} & = & \vec{S}_\chi \cdot \left(  \vec{S}_N \times 
\vec{v}^{\perp} \right) \nonumber \\
\mathcal{O}_{15} & = & - \left(\vec{S}_\chi \cdot \frac{\vec{q}}{m_N}\right)
\left[  \vec{S}_N \times \left(
\vec{v}^{\perp} \right) \cdot  \frac{\vec{q}}{m_N} \right] \quad, 
\end{eqnarray}
where $m_N$ is the nucleon mass, $\vec{S}_\chi$ and $\vec{S}_N$  are  the WIMP
and the nucleon spins, $\vec{q}$ is 
momentum transfer in the collision,  and $\vec{v}^{\perp} \equiv \vec{v} -\vec{q} 
(\vec{v} \cdot \vec{q})/ q^2 = \vec{v} + \vec{q}  /(2\mu_T)$
is the transverse relative velocity. The last equality follows from energy conservation
and $\mu_T\equiv (m_\chi m_T )/ (m_\chi+m_T )$ is the reduced mass between a WIMP of
mass $m_\chi$ and a target nucleus of mass $m_T$.
Operators $\mathcal{O}_{12}$ and $\mathcal{O}_{15}$ can appear only for 
mediators with spin greater than one.
Since the typical energy transfer in 
WIMP-nucleus collision is much lower than the nuclear binding energy, and the 
collision is essentially nonrelativistic, 
the differential elastic cross section can be naturally organized so that nuclear and
particle physics factorize \cite{Anand:2013yka} as follows:
\begin{eqnarray} \label{eq:csec}
\frac{d \sigma_N}{dE_R} (q,v) &=& 
\frac{2 m_T }{v^2} 
\sum_{k} 
R_k \left( \vec{v}_T^{\perp 2}, {\vec{q}^{\,2} \over m_N^2} \right)
 W_k^{00}( \vec{q}^{\,2})\\
&=& 
\frac{2 m_T W_M^{00}(0) }{v^2} 
\sum_{k} 
R_k \left( \vec{v}_T^{\perp 2}, {\vec{q}^{\,2} \over m_N^2} \right)
 \frac{W_k^{00}( \vec{q}^{\,2})}{W_M^{00}(0)}  \nonumber 
\end{eqnarray}
where $E_R=\vec{q}^{\;2}/(2 m_T)$ is the nucleus recoil energy,  $m_T$ is the mass of the target 
nucleus, the $R_k$'s are the WIMP response functions, which depend parametrically on 
the operator coupling coefficients $\left\{c_i \right\}$, 
and the $W_k^{00}$ are the corresponding nuclear response functions.
These response functions generalize the standard form factor, which reflects the finite size of the nucleus, by taking into account the velocities of the nucleons.
The ``$00$" superscript indicates the isoscalar-isoscalar combination, as in Ref.~\cite{Anand:2013yka}. For spin-zero nuclei, three
response functions appear, $k=M$, $\Phi''$, or $M\Phi''$ using the notation of Ref.~\cite{Anand:2013yka}. 
If only $c_1$, the coupling of the SI operator $\mathcal{O}_1$, is different from zero, then
only  $R_M = c_1^2$ appears. 
In this case Eq.~(\ref{eq:csec}) reduces to the standard SI result:
\begin{equation}
\label{eq:csecStandard1}
 \frac{d \sigma_N}{dE_R} (q,v) =
\frac{2 m_T c_1^2 }{v^2} 
W_M^{00}( \vec{q}^{\,2})
 = \frac{A^2 \sigma_1}{\mu_N^2}
\frac{m_T }{2v^2} 
 \frac{W_M^{00}( \vec{q}^{\,2})}{W_M^{00}(0)}  \, ,
\end{equation}
where we have defined the 
 WIMP-nucleon cross section
\begin{equation}
\label{eq:sigma1}
 \sigma_{1} \equiv
 c_1^2 \mu_N^2 \frac{4 W^{00}_M(0)}{A^2}  \, ,
\end{equation}
with $ \mu_N$ the WIMP-nucleon 
reduced mass and $A$ the mass number. The  normalized response function,
$W_M^{00}(\vec{q}^{\,2}) / W_M^{00}(0) $, corresponds to the square 
of  the form factor that
is often  parametrized using the  Helm form factor \cite{bib:halo1}.

When a more general interaction is considered, the response functions $R_k$'s 
can be dependent on the momentum transfer and on the  relative velocity of the incoming particles.
One can classify the various contributions to the differential cross section 
according to the powers of $\vec{q}^{\; 2}=2m_T E_R$
and $\vec{v}^{\perp 2}$ that appear in the WIMP response functions $R_k$. 
Equations~(37) and (38) in Ref.~\cite{Anand:2013yka} show the contributions to the
elastic differential cross section in Eq.~(\ref{eq:csec}). These
contributions have the following powers of
 $\vec{q}^{\; 2}$ and $\vec{v}^{\perp 2}$:
 \begin{itemize}
  \item 
  the WIMP response function $R_{M}^{00}$, which multiples  the nuclear response function $W^{00}_{M}$, has four
terms,
proportional to 1, $\vec{q}^{\; 2}$, $\vec{v}^{\perp 2}$, and $\vec{q}^{\; 2} \cdot \vec{v}^{\perp 2}$;
\item
the WIMP response function $R_{\Phi''}^{00}$, which multiples  the nuclear response function $W_{\Phi''}^{00}$, has three
terms, proportional to $\vec{q}^{\; 2}$, $\vec{q}^{\; 4}$, and $\vec{q}^{\; 6}$;
\item
finally, the WIMP response function $R_{M\Phi''}^{00}$, which multiples the nuclear response function $W^{00}_{M\Phi''}$, has 
two contributions  proportional to $\vec{q}^{\; 2}$, and $\vec{q}^{\; 4}$.
 \end{itemize}
Since in the kinematic regime of interest higher powers of $\vec{q}^{\; 2}$ are expected to be subdominant, we choose to
leave out the term proportional to $\vec{q}^{\; 6}$. The EFT expansion in Eq.~(4) is left with
eight contributions that differ because they have different powers of $\vec{q}^{\; 2}$ or $\vec{v}^{\perp 2}$ or
different nuclear response functions.

If we include the possibility that the interaction mediator 
could be much lighter than the  momentum transfer and,
therefore, that the differential cross section could contain an additional factor 
proportional to $(\Lambda/q)^4$ with $\Lambda$ a momentum scale, 
we find eight additional possibilities for a total of
16 possible combinations of powers of  $\vec{q}^{\; 2}$
or  $\vec{v}^{\perp 2}$ and nuclear responses. 
A similar classification of the possible interactions have been proposed in Ref. \cite{bib:guo}. Reference \cite{bib:guo}, however, considers also terms
proportional to $\vec{v}^{\perp 4}$, but such terms do not arise in EFT [see Eq.~(38) in Ref.~\cite{Anand:2013yka}], and does not take into account that additional operators could probe different form factors.
Given a specific theoretical model, where the ratios between all the couplings $c_i$  are given, we could make an exclusion
curve as a function of an overall scale of the interaction. In the standard approach only $c_1$ is assumed different from zero. 
In the same spirit of probing a single coupling at the time, 
this work shows results for the cases when only one coefficient in the
expansion in Eq.~(\ref{eq:OperatorSum}) is different from zero.
Table \ref{tab:ops} lists the 12 remaining terms of the expansion: the four terms that
multiply the mixed nuclear response function $M\Phi''$ have not been considered, since they appear when 
at least two  $c_i$ are different from zero.
Note that, in principle, the power-counting 
classification and the implied relative 
importance of the different contributions could be modified by QCD effects; see for instance the
chiral EFT in Ref. ~\cite{Hoferichter:2016nvd}, or by fine-tuning the $c_i$  parameters of the nucleus-WIMP interaction. 
Each of the 12 terms of the EFT expansion leads to a term in the differential cross section 
\begin{eqnarray}
\label{eq:csecStandard3}
\frac{d \sigma_N}{dE_R} (q,v) &=& 2 c_i^2 d_i
\frac{m_T }{v^2} \left( \frac{q}{q_\mathrm{ref}} \right)^{2\alpha} \left( \frac{v^\perp}{v_\mathrm{ref}} \right)^{2\beta}
W_k^{00}( \vec{q}^{\,2})  \\
\label{eq:csecStandard4}
 &=&  \frac{A^2 \sigma_i}{\mu_N^2}
\frac{m_T }{2v^2} \left( \frac{q}{q_\mathrm{ref}} \right)^{2\alpha} \left( \frac{v^\perp}{v_\mathrm{ref}} \right)^{2\beta}
 \frac{W_k^{00}( \vec{q}^{\,2})}{W_M^{00}(0)}\,,
\end{eqnarray}
where $\alpha=0$, 1 or 2 and $\beta=0$ or 1, $d_i$ are dimensionless coefficients, which are explicitly given in the last column of Table~\ref{tab:ops} and $k$ labels the nuclear response function. In analogy with Eqs. (\ref{eq:csecStandard1})  and  (\ref{eq:sigma1})
we have also defined a cross section $\sigma_i\equiv c^2_i d_i ( \sigma_1/c_1^2) $ for each term and we have introduced $ q_{\rm  ref}$ and $v_{\rm ref}$, typical momentum transfer and velocity in a direct dark matter phenomenology so that $\sigma_i$
has the dimension of a cross section. Specific theoretical models fix the values of 
$\sigma_i/(q_{\rm  ref}^{2\alpha} v_{\rm ref}^{2\beta} )$.
 A different choice would
scale $\sigma_i \to \sigma_i (q_{\rm  ref}'/q_{\rm  ref})^{2\alpha}
(v_{\rm ref}'/v_{\rm ref})^{2\beta}$.
We present our results using $ q_{\rm  ref}= 100$~MeV/c and $v_{\rm ref}= v_0 = 220$~km/s, the standard halo local velocity. The nuclear response functions
$W^{00}_M$ and $W^{00}_{\Phi''}$ for $^{40}$Ar
have been taken from Ref. \cite{Catena:2015uha}.

\begin{table} 
\centering
\begin{tabular}{|c|c|c|c|} 
\hline
Operator &      $ R_k $        & Nuclear &  $d_i$ \\
Coupling & Expansion  & Response &  \\
\hline
\hline
$c_1^2$ & $ 1$ & $W^{00}_M$ & 1\\
$c_{11}^2$ &  $ \left(\frac{q}{q_\mathrm{ref}}\right)^2  $ &  $W^{00}_M$ & 
 $\frac{j_\chi (j_\chi +1)}{3} \left(\frac{q_\mathrm{ref}}{m_N}\right)^2  $ \\
$c_{8}^2$  & $ \left(\frac{v^{\perp}}{v_\mathrm{ref}}\right)^2$ &  $W^{00}_M$  & 
 $\frac{j_\chi (j_\chi +1)}{3}  v_\mathrm{ref}^2  $ \\
 $c_5^2$ & 
$ \left(\frac{q}{q_\mathrm{ref}}\right)^2 \left(\frac{v^{\perp}}{v_\mathrm{ref}}\right)^2$ 
&  $W^{00}_M$ &  
 $\frac{j_\chi (j_\chi +1)}{3} \left(\frac{q_\mathrm{ref}}{m_N}\right)^2 v_\mathrm{ref}^2  $ \\
\hline
 $c_{12}^2$ & $ \left(\frac{q}{q_\mathrm{ref}}\right)^2$  & $W^{00}_{\Phi''}$  &
 $\frac{j_\chi (j_\chi +1)}{12} \left(\frac{q_\mathrm{ref}}{m_N}\right)^2  $ \\
 $c_3^2$& $ \left(\frac{q}{q_\mathrm{ref}}\right)^4$ & $W^{00}_{\Phi''}$ &
 $\frac{1}{4} \left(\frac{q_\mathrm{ref}}{m_N}\right)^4  $
 \\
\hline
$c_1^{*2}$ & $ \left(\frac{q_\mathrm{ref}}{q}\right)^4$ & $W^{00}_M$ & 
\\
$c_{11}^{*2}$ &  $  \left(\frac{q_\mathrm{ref}}{q}\right)^2 $ &  $W^{00}_M$ & \\
$c_{8}^{*2}$  & $  \left(\frac{q_\mathrm{ref}}{q}\right)^4 \left(\frac{v^\perp}{v_\mathrm{ref}}\right)^2 $ &  $W^{00}_M$  &\\
$c_5^{*2}$ & $  \left(\frac{q_\mathrm{ref}}{q}\right)^2 \left(\frac{v^\perp}{v_\mathrm{ref}}\right)^2$ &  $W^{00}_M$ &\\
\hline
$c_{12}^{*2}$ & $  \left(\frac{q_\mathrm{ref}}{q}\right)^2$  & $W^{00}_{\Phi''}$  &\\
$c_3^{*2}$
& $ 1$ & $W^{00}_{\Phi''}$ &\\
\hline
\hline
\end{tabular} 
\caption{
List of addition powers of $q$ and $v^{\perp}$ relative to the SI scalar operator in
the nonrelativistic EFT expansion in Eq. (\ref{eq:OperatorSum})  of the differential cross 
section in Eq. (\ref{eq:csec}),
when only operators contributing to spin-zero nuclei are considered and only one of the 
couplings $c_i$  in Eq. (\ref{eq:OperatorSum}) is different from zero.
The first column shows the $c_i$'s, following the notation of Ref. \cite{Anand:2013yka},
whereas the second column shows the corresponding powers of $q$ and $v^{\perp}$ appearing in the 
WIMP response functions $R_k$ and finally the third column lists the corresponding nuclear 
response functions associated to the operator. The fourth column shows the dimensionless coefficient $d_i$ that appears
in Eq. (\ref{eq:csecStandard3}), where $m_N$ is the nucleon mass, $j_\chi$ is the WIMP spin,
and $v_\mathrm{ref}$ is relative to the speed of light.
The star $*$ denotes cases with a light mediator with
propagator $(\Lambda/q)^4$; the relations between the 
$\sigma_i^*$'s and $c_i^*$ are the same as the case of the heavy mediator, but the 
$c^*_i$ change with $q_\mathrm{ref}$ as
$c^*_i=c_i (\Lambda/q_\mathrm{ref})^2$ given the operator combination of
Eq.~(\ref{eq:OperatorSum}).} \label{tab:ops}
\end{table}

The total interaction rate $R$ is obtained from Eq. (\ref{eq:csecStandard4}) by integrating 
over the recoil energy $E_R$ in the experimental window and over the WIMP velocities
\begin{equation} \label{eq:rate}
R = N_T \frac{\rho}{m_\chi} \int dE_R \int d^3v \frac{d \sigma_N}{dE_R} 
(E_R, v)  v f(v)\,,
\end{equation} 
where $N_T$ is the number of target nuclei, $\rho=0.3$ GeV/ (c$^2$ cm$^{3})$ is the 
local dark matter  density, and  $f(v)$ is a Maxwellian velocity
distribution \cite{bib:halo1} with a cutoff $v_{\rm esc} = 544$ km/s 
 \cite{bib:halo2,bib:halo3} and
velocities $v_0= 220$~km/s and $v_E =232$ km/s \cite{bib:halo4}. 

\begin{figure*}[!ht]
\centering
\includegraphics[width=0.9\textwidth]{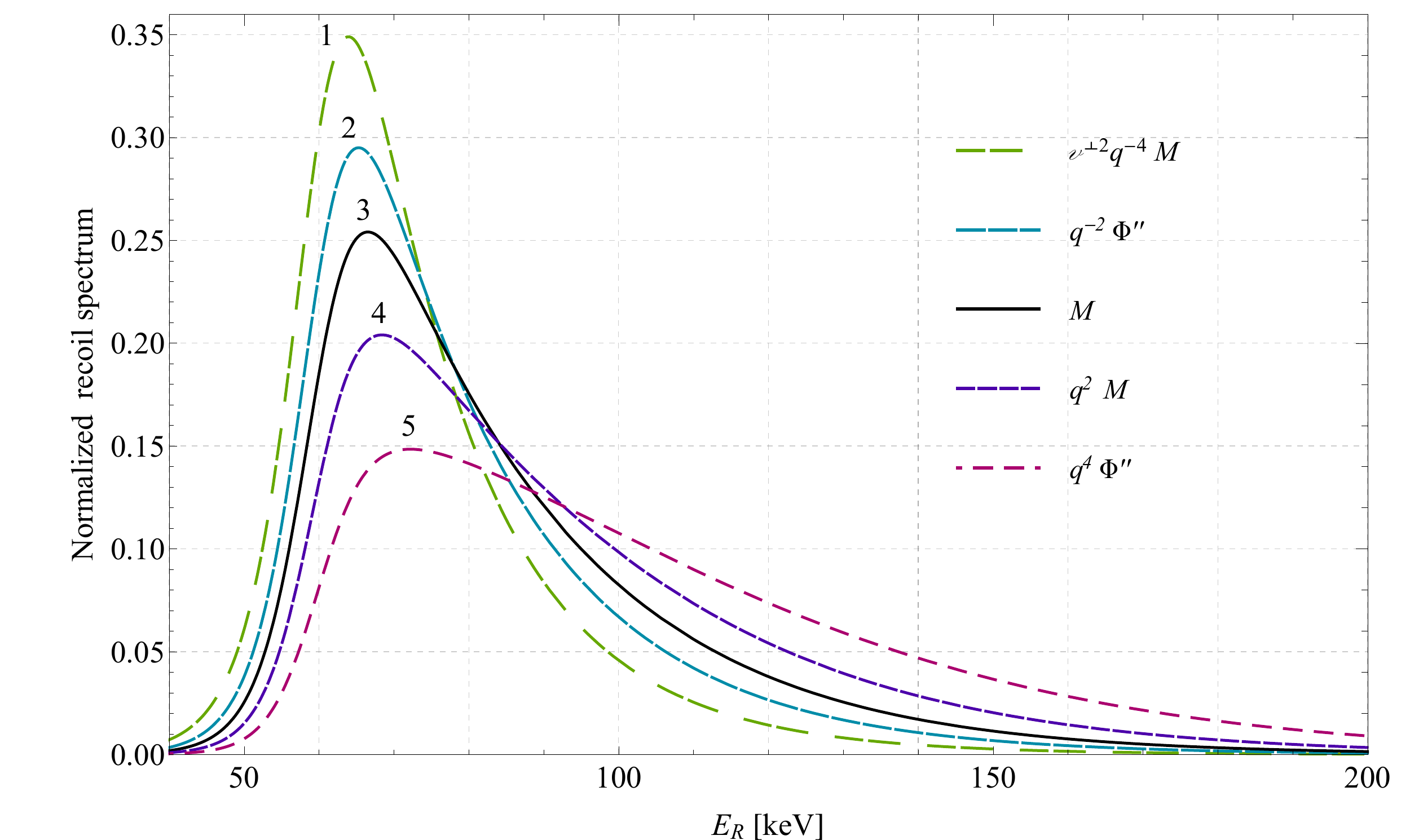}
\caption{Expected recoil-energy spectra of argon nuclei in DarkSide-50 from the interaction of
100~$\mathrm{MeV/c^2}$ WIMPs with the SHM velocity distribution for five different EFT operators. Spectra include the acceptance of
the detector and are arbitrary normalized. Curve labeled (3) shows the standard spectrum
corresponding to the SI operator, i.e., the form factor $M$ in the adopted notation. The 
other four curves correspond to (1) the nuclear response function $M$ times the
the factor $v^{\perp^2} q^{-4} $, (2)  $\Phi''$ times  $q^{-2}$, (4) $M$ times the
factor $q^{2}$, and (5) $\Phi''$ times  $q^{4}$.\\
}
\label{fig:spectra}
\end{figure*}

Since the DarkSide-50 experiment has not detected any WIMP event,
limits for each of the 12 cross sections $\sigma_i$ are given as a function of the 
WIMP mass $M_\chi$. Figure~\ref{fig:spectra} shows the normalized shape of the recoil energy
for five selected operators in an argon detector with the acceptance of
DarkSide-50~\cite{Agnes:2018fwg}. The solid curve (number 3) corresponds
to the standard SI operator.
The other four curves are examples which give the most extreme results in terms of the final WIMP-nucleus cross-section exclusion limits for each of the two response functions $\Phi''$ and $M$.
Given enough WIMP events the recoil spectrum should make it possible to distinguish between different interaction models. A statistical analysis that takes into account the different expected recoil spectra gives stronger exclusion curves if background is present; this is not our case, since the DarkSide-50 experiment has a total expected background after the selection of only about 0.1 events.

In the experimental realizations, the rate in Eq. (\ref{eq:rate}) is convolved 
with detector resolution and the energy scale must be rescaled according to the 
relation
\begin{equation}
Q(E_R) = LY \times E_R \times \mathcal{L}_{\rm eff} (E_R),
\end{equation}
where $Q(E_R)$ is the energy estimator, $LY$ is the light yield in photoelectrons (PEs) per keV and $\mathcal{L}_{\rm eff} (E_R)$ is the nuclear-recoil quenching. In this new 
variable Eq.~(\ref{eq:csec}) becomes
\begin{equation} 
 \frac{d \sigma_N}{dE_R} (q,v) \rightarrow \frac{d \sigma_N}{dE_R}  
\frac{dE_R}{dQ} \otimes \mathcal{R}(Q),
\end{equation}
where $\mathcal{R}$ is the resolution function and $\otimes$ denotes the 
convolution product. The calibration of the energy scale for nuclear recoils and the experimental resolution are  briefly described in the next section.

\section{EFT limits in DarkSide-50 experiment} \label{sec:ds50}

The DarkSide-50 experiment, located at Laboratori Nazionali del Gran Sasso 
(LNGS), following the results of its predecessor DarkSide-10~\cite{bib:ds10}, 
searches for nuclear recoils (NRs) induced by WIMP scattering with a 
liquid argon double-phase time projection chamber (LAr-TPC), surrounded by a 
spherical liquid scintillator veto (LSV) located in the center of a cylindrical 
water Cherenkov veto. The active veto detectors are used for rejecting the 
coincidences in the LAr-TPC induced by cosmic and material radiation (see, for 
details,~\cite{bib:veto,bib:ds50_50d,bib:ds50_70d,Agnes:2015qyz,Agnes:2016nis,Agnes:2017dfc,Agnes:2017grb}).
Two arrays of 19 Photo Multipliers each of 3'', facing from the top and the bottom the liquid argon 
active volume ($\sim46.4$ kg), detect the primary scintillation light (whose signal is called S1) and 
the gas scintillation from drifted ionization electrons (whose signal is called S2).
LAr intrinsic scintillation characteristics allow us to reject electron recoils (ERs), 
essentially beta and gamma events from background, at the level of 
$1.5\times 10^7$ or even better~\cite{bib:ds50_50d}. The particle identification is based
on  the fraction of S1 detected in the first 90 ns from the pulse start time ($f_{90}$ parameter).

\begin{table*} 
\begin{center}
\begin{tabular}{|c|c|c|}
\hline
   & \multicolumn{2}{|c|}{ $\sigma_i$  ($\mathrm{cm}^2$) }  \\
      \hline
 Model & $M_{\chi}$ = 100 $\mathrm{GeV/c^2}$ &
     $M_{\chi}$ = 1000 $\mathrm{GeV/c^2}$ \\
   \hline 
    $q^4 \Phi''$     & 2.3 $\times 10^{-42}$  & 6.0 $\times 10^{-42}$   \\
    $q^2 \Phi''$     & 1.6 $\times 10^{-42}$  & 4.9 $\times 10^{-42}$   \\
    $\Phi''$        & 1.0 $\times 10^{-42}$  & 3.5 $\times 10^{-42}$   \\
    $q^{-2} \Phi''$  & 6.2 $\times 10^{-43}$  & 2.3 $\times 10^{-42}$   \\
    $q^2$$M$        & 1.8 $\times 10^{-44}$  & 5.5 $\times 10^{-44}$   \\
    $M$            & 1.1 $\times 10^{-44}$  & 3.8 $\times 10^{-44}$   \\
    $v^{\perp^2} q^2$$M$   & 1.2 $\times 10^{-44}$  & 3.5 $\times 10^{-44}$    \\
    $q^{-2}$$M$    & 6.6 $\times 10^{-45}$  & 2.5 $\times 10^{-44}$   \\
    $v^{\perp^2}$$M$        & 7.4 $\times 10^{-45}$  & 2.5 $\times 10^{-44}$   \\
    $v^{\perp^2} q^{-2}$$M$ & 4.3 $\times 10^{-45}$  & 1.6 $\times 10^{-44}$   \\
    $q^{-4}$$M$    & 3.7 $\times 10^{-45}$  & 1.5 $\times 10^{-44}$   \\
    $v^{\perp^2} q^{-4}$$M$ & 2.4 $\times 10^{-45}$  & 8.9 $\times 10^{-45}$   \\
   \hline
\end{tabular} 
\caption{Values of the cross section parameters $\sigma_i$ 
for the 12 EFT terms as defined in Eq.~(\ref{eq:csecStandard4})
excluded at the $90\%$ C.L. for two values of the WIMP mass.
\label{tab:sinopt} }
\end{center}
\end{table*}

The DarkSide-50 experiment took data in two campaigns: 
first, the atmospheric argon campaign, in which the main features of the 
detector have been understood and tested~\cite{bib:ds50_50d}; second, the 
underground depleted argon (UAr) campaign in which the predicted characteristics 
have been confirmed and the impressive reduction of the $^{39}$Ar isotope has been 
proven~\cite{bib:ds50_70d}.

UAr was extracted in Colorado gas plants, purified at Fermilab and shipped to 
LNGS, during an intense cooperation of many years \cite{bib:uar1}. The 
$^{39}$Ar activity of UAr is a factor $(1.4\pm0.2) \times 10^3$ lower than the 
atmospheric argon one, corresponding to an activity of $(0.73\pm0.11)$ mBq/kg 
\cite{bib:ds50_70d}. 

The TPC response calibration is performed with neutron and gamma sources and with gaseous ${^{83m}}$Kr injected into the target volume~\cite{Response}. The S1 scintillation efficiency of nuclear recoils was measured with test beam experiments, namely SCENE~\cite{SCENE} and ARIS~\cite{ARIS}, and cross-calibrated with AmBe and AmC neutron sources 
in DarkSide-50~\cite{Agnes:2018ves}.
The analysis uses both S1 and S2. S1 gives information on the nature of the event  and
is the main energy variable. However, a combination of S1 and S2 
gives an energy variable with better resolution and linearity, since
the deposited energy is shared between scintillation and ionization.
In addition, S2 determines the position and
rejects multiple scatter events.
Reference~\cite{bib:ds50_50d} describes the procedure to calibrate the 
nuclear-recoil energy scale from the scintillation
signal using the PE yield for nuclear recoils of known energy
measured in the SCENE experiment~\cite{SCENE}.
In summary, SCENE measures
the ratio between the PE yield from NR at 200 V/cm and that from ${^{83m}}$Kr at
zero field. The DarkSide-50 zero-field PE yield for ${^{83m}}$Kr
(8.0$\pm$0.2~PE/keV~\cite{Agnes:2018fwg} measured at the peak energy of 41.5~keV) then gives the NR PE yield vs. S1. We assume constant NR PE yield above the highest
SCENE-measured energy, $\sim57.3$ keV$_{\mathrm{nr}}$.
Monte Carlo simulations estimate that the 
overall S1 light collection efficiency, averaged on the entire volume,
is about $\sim$~16\%.
The analysis of the DarkSide-50 data is performed in blind mode as explained in Ref.~\cite{Agnes:2018fwg}.
The expected background events can be classified into three categories: surface events, neutrons (cosmogenic and radiogenic), and ERs. Surface events are mostly rejected with 
fiducialization of the active volume, neutrons are efficiently suppressed with the LSV, and ERs are rejected with high efficiency using the $f_{90}$ parameter.  The LSV, whose estimated efficiency is 0.9964$\pm$0.0004, identified 4 neutron candidates. After the LSV cut, the dominant background comes from ERs (0.08$\pm$0.04 surviving events). The $f_{90}$ acceptance requires a relatively large 
nuclear-recoil threshold energy. The final acceptance is 60.9\%, with a threshold energy $\gtrsim$50~keV$_{\mathrm{nr}}$ (see Fig. 10 of Ref.~\cite{Agnes:2018fwg}) and the fiducial mass corresponds to 36.9$\pm$0.6 kg. 
The number of expected surviving background events for the entire statistics, which corresponds to (16660 $\pm$ 270) kg d exposure, is $0.09\pm0.04$ (for a detailed summary  see Table V of Ref.~\cite{Agnes:2018fwg}).
After the data unblinding, no events were observed in the defined WIMP search region, as shown in Fig. 11 (right) of Ref.~\cite{Agnes:2018fwg}.
The lack of observed events is consistent with up to 2.3 WIMP-nucleon scatters expected at 90\% C.L. and so can be used to draw 90\% C.L. exclusion curves for the $\sigma_i$ cross sections in terms of the 12 realizations enumerated in Table~\ref{tab:ops},  using a simple cut and counts statistical technique.

\begin{figure*}[!ht]
\centering
\includegraphics[width=.9\textwidth]{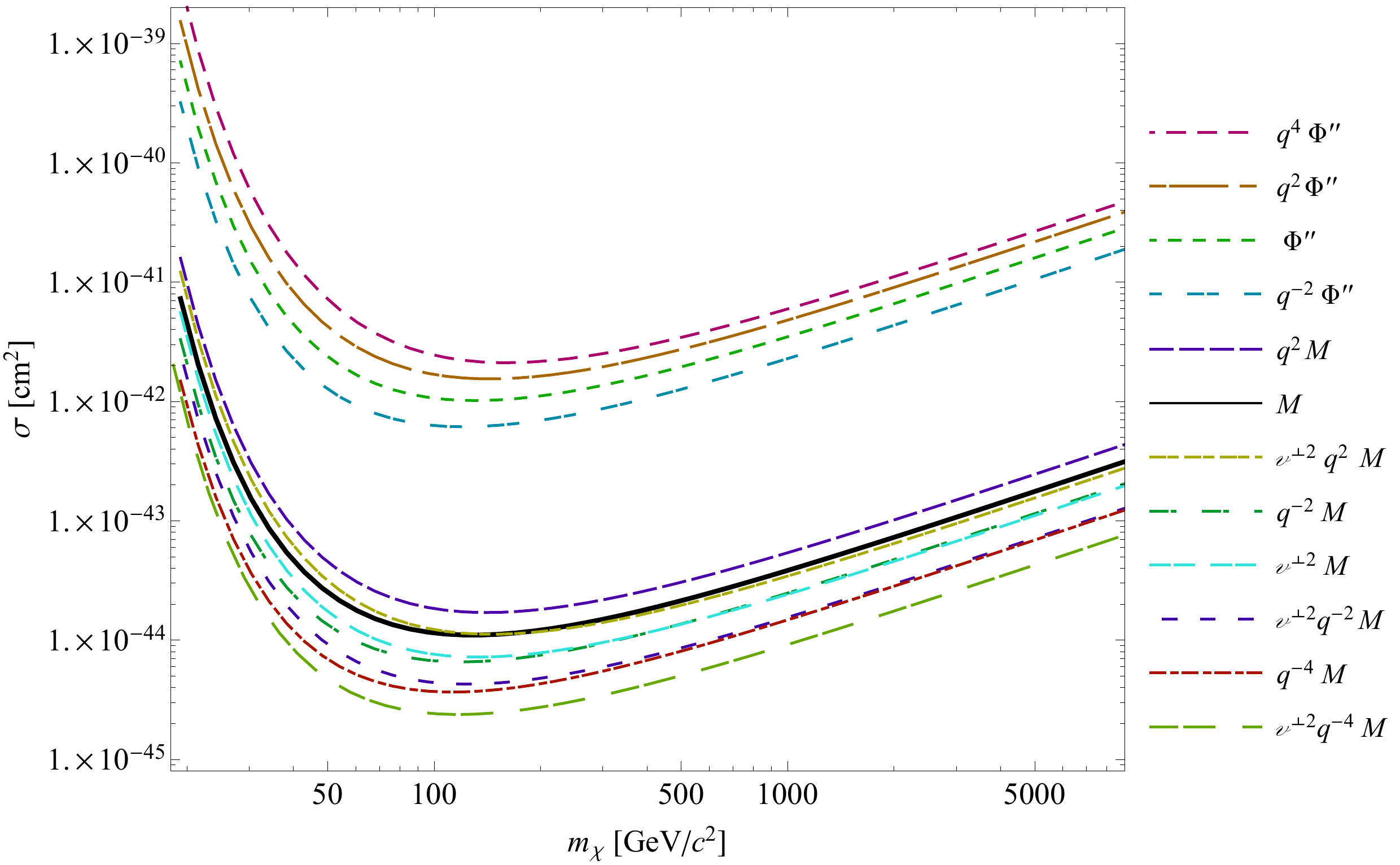}
\caption{DarkSide-50 90\% C.L. exclusion curves on the cross section parameter $\sigma_i$ 
for the 12 EFT terms as defined by Eq. (\ref{eq:csecStandard4}). 
Going from top to bottom, we see a group of four curves that correspond to the 
nuclear response function $\Phi''$ times $q^4$, $q^2$, 1, or $q^{-2}$; then a group
of eight curves corresponding to the nuclear response function $M$ times
$q^2$, 1, $q^2 v^{\perp 2}$, $q^{-2}$, $v^{\perp 2}$, $q^{-2} v^{\perp 2}$, $q^{-4}$,
or $q^{-4} v^{\perp 2}$. The 
solid black curve represents the standard spin-independent limit that corresponds to the current limit published in Ref.~\cite{Agnes:2018fwg}. }
\label{fig:ds50}
\end{figure*}

Note that a general  relativistic WIMP-nucleon interaction can be expanded
in the  nonrelativistic EFT operator base of Eq.~(\ref{eq:OperatorSum}) resulting in 
a linear combination of the terms listed in Table~\ref{tab:ops}. However, the 
corresponding 90\% C.L. exclusion curve cannot be immediately deduced by the 
individual curves for each NR operator.

There are two  groups of curves in Fig.~\ref{fig:ds50}: the eight curves at the bottom correspond 
to the standard spin-independent 
coherent response function $M$, and the four curves at the top correspond to the form factor $\Phi''$ and give much weaker limits. This last form factor is related to spin-orbit coupling mainly of
the two unpaired neutrons and the two proton holes in $^{40}$Ar and it is therefore about a factor
$(4/40)^2$ smaller than $M$. Within each group, the operator
proportional to the smaller power of $q$ gets the stronger limit, since the expected rates are higher when lower recoil energies have larger
weight. Table~\ref{tab:sinopt} shows the 90\% C.L. limits for the 12 cross sections  for  WIMPs of mass of 100 $\mathrm{GeV/c^2}$ and 1000 $\mathrm{GeV/c^2}$.

\section{Conclusions} \label{sec:conclusions}
We have reanalyzed the latest DarkSide-50 results with a total exposure of $(16660 \pm 270)$ kg d 
in terms of the 12 leading effective operators naturally appearing in a 
nonrelativistic expansion.
This extended set of operators leads to 90\% C.L. upper limits on the effective couplings that parametrize
the WIMP-nucleon interaction. These couplings, one of which is the coherent SI standard interaction span many orders of magnitude.  Figure~\ref{fig:ds50} shows the experimental constraints as a function of the WIMP mass and in Table~\ref{tab:sinopt} the corresponding numerical values for WIMPs of masses of 100 $\mathrm{GeV/c^2}$ and 1000 $\mathrm{GeV/c^2}$ are highlighted.
For instance, for the interaction parametrized only by the operator leading
to the nuclear response  function $M$ times $q^{-4}v^{\bot2}$, the DarkSide-50 data yield
a 90\% confidence limit on the corresponding cross section, as defined in Eq.~(\ref{eq:csecStandard4}), of 
$2.4 \times 10^{-45} \,\mathrm{cm}^2$ (8.9 $\times 10^{-45} \, \mathrm{cm}^2$) for a WIMP mass of
100 (1000) GeV/c$^2$,  which is a factor about
five more stringent than the standard SI limit. On the contrary, for the interaction parametrized by  the $\Phi''$ nuclear function times 
$q^4$, the limit on the corresponding cross section is only
$2.3 \times 10^{-42} \, \mathrm{cm}^2$
($6.0 \times 10^{-42} \, \mathrm{cm}^2$) for a 100 (1000) GeV/c$^2$, more than 2 orders of magnitude larger than the standard SI limit.
Different operators also predict different WIMP recoil spectra, as shown in Fig.~\ref{fig:spectra}. Thus, different interaction models could be tested if enough WIMP events will be detected in the future.
Moreover, the relative importance of the different EFT operators depends on the target nuclei that can have very
different response functions. One should be prudent when comparing 
limits and/or signals from experiments with different targets under the assumption of the simplest interaction model,
the SI scalar cross section. The complementarity of experiments using different targets could be crucial 
for probing the full parameter space.

\section*{ACKNOWLEDGMENTS}

The DarkSide Collaboration offers its profound gratitude
to the LNGS and its staff for their invaluable technical and
logistical support. We also thank the Fermilab Particle
Physics, Scientific, and Core Computing Divisions.
Construction and operation of the DarkSide-50 detector
was supported by the U.S. National Science Foundation
(NSF) (Grants No. PHY-0919363, No. PHY-1004072,
No. PHY-1004054, No. PHY-1242585, No. PHY-1314483,
No. PHY-1314501, No. PHY-1314507, No. PHY-1352795,
No. PHY-1622415, and associated collaborative grants
No. PHY-1211308 and No. PHY-1455351), the Italian
Istituto Nazionale di Fisica Nucleare, the U.S. Department of Energy (Contracts No. DE-FG02-91ER40671, No. DEAC02-07CH11359, and No. DE-AC05-76RL01830), the
Russian Science Foundation (Grant No. 18-72-00211), the
Polish NCN (Grant No. UMO-2014/15/B/ST2/02561) and
the Foundation for Polish Science (Grant No. Team2016-2/
17). We also acknowledge financial support from the French
Institut National de Physique Nucléaire et de Physique des 
Particules (IN2P3), the UnivEarthS Labex program of
Sorbonne Paris Cité (Grants No. ANR-10-LABX-0023
and No. ANR-11-IDEX-0005-02), and from the S\~ao Paulo
Research Foundation (FAPESP) (Grant No. 2016/09084-0).
Isotopes used in this research were supplied by the United
States Department of Energy Office of Science by the Isotope
Program in the Office of Nuclear Physics



\begin{thebibliography}{99}


\bibitem{bib:dm1} S. M. Faber and J. S. Gallagher, Annu. Rev. Astro.
Astrophys. 17, 135 (1979).

\bibitem{bib:dm2} D. N. Spergel, Phys. Rev. D 37, 1353 (1988).

\bibitem{bib:dm3} D. Clowe et al., Ap. J. 648, L109 (2006).

\bibitem{bib:dm4}J. L. Feng, Annu. Rev. Astro. Astrophys. 48, 495
(2010).

\bibitem{bib:eft1} A. L. Fitzpatrick, W. Haxton, E. Katz, N. Lubbers, and
Y. Xu, ArXiv e-prints (2012), arXiv:1211.2818 [hep-ph].

\bibitem{Fitzpatrick:2012ix} 
  A.~L.~Fitzpatrick, W.~Haxton, E.~Katz, N.~Lubbers and Y.~Xu,
  JCAP {\bf 1302}, 004 (2013)
  doi:10.1088/1475-7516/2013/02/004
  [arXiv:1203.3542 [hep-ph]].
  
\bibitem{Anand:2013yka} 
  N.~Anand, A.~L.~Fitzpatrick and W.~C.~Haxton,
  Phys.\ Rev.\ C {\bf 89}, no. 6, 065501 (2014)
  doi:10.1103/PhysRevC.89.065501
  [arXiv:1308.6288 [hep-ph]].

\bibitem{bib:halo1} J.D. Lewin and P. Smith, Astropart. Phys. 6, 87 (1996).

\bibitem{bib:halo2} C. Savage, K. Freese, and P. Gondolo, Phys. Rev. D 74, 
043531 (2006).

\bibitem{bib:halo3} M.C. Smith et al., Mon. Not. Roy. Astron. Soc. 379, 755 
(2007).

\bibitem{bib:halo4} C. Savage, G. Gelmini, P. Gondolo, and K. Freese, JCAP 0904, 
010 (2009).

\bibitem{bib:dama} R. Bernabei et al., Eur. Phys. J. C 67, 39 (2010).
\bibitem{Bernabei:2018yyw} 
  R.~Bernabei {\it et al.},
  Universe {\bf 4}, no. 11, 116 (2018)
  doi:10.3390/universe4110116
  [arXiv:1805.10486 [hep-ex]].



\bibitem{Agnese:2013rvf} 
  R.~Agnese {\it et al.} [CDMS Collaboration],
  Phys.\ Rev.\ Lett.\  {\bf 111}, no. 25, 251301 (2013)
  doi:10.1103/PhysRevLett.111.251301
  [arXiv:1304.4279 [hep-ex]].

\bibitem{bib:xenon} E. Aprile et al. (The XENON100 Collaboration),
Phys. Rev. Lett. 109, 181301 (2012).


\bibitem{Akerib:2017kat} 
  D.~S.~Akerib {\it et al.} [LUX Collaboration],
  Phys.\ Rev.\ Lett.\  {\bf 118}, no. 25, 251302 (2017)
  doi:10.1103/PhysRevLett.118.251302
  [arXiv:1705.03380 [astro-ph.CO]].


\bibitem{Cui:2017nnn} 
  X.~Cui {\it et al.} [PandaX-II Collaboration],
  Phys.\ Rev.\ Lett.\  {\bf 119}, no. 18, 181302 (2017)
  doi:10.1103/PhysRevLett.119.181302
  [arXiv:1708.06917 [astro-ph.CO]].


\bibitem{Aprile:2018dbl} 
  E.~Aprile {\it et al.} [XENON Collaboration],
  Phys.\ Rev.\ Lett.\  {\bf 121}, no. 11, 111302 (2018)
  doi:10.1103/PhysRevLett.121.111302
  [arXiv:1805.12562 [astro-ph.CO]].



\bibitem{bib:nsi1}
  K.~Schneck {\it et al.} [SuperCDMS Collaboration],
  Phys.\ Rev.\ D {\bf 91} (2015) no.9,  092004
  doi:10.1103/PhysRevD.91.092004
  [arXiv:1503.03379 [astro-ph.CO]].

\bibitem{bib:nsi2}
  V.~Gluscevic, M.~I.~Gresham, S.~D.~McDermott, A.~H.~G.~Peter and K.~M.~Zurek,
  JCAP {\bf 1512} (2015) no.12,  057
  doi:10.1088/1475-7516/2015/12/057
  [arXiv:1506.04454 [hep-ph]].
  
\bibitem{bib:nsi3}
  M.~Cirelli, E.~Del Nobile and P.~Panci,
  JCAP {\bf 1310} (2013) 019
  doi:10.1088/1475-7516/2013/10/019
  [arXiv:1307.5955 [hep-ph]].
  
\bibitem{Aprile:2017aas} 
  E.~Aprile {\it et al.} [XENON Collaboration],
  Phys.\ Rev.\ D {\bf 96}, no. 4, 042004 (2017)
  doi:10.1103/PhysRevD.96.042004
  [arXiv:1705.02614 [astro-ph.CO]].

\bibitem{Xia:2018qgs} 
  J.~Xia {\it et al.} [PandaX-II Collaboration],
  arXiv:1807.01936 [hep-ex].

\bibitem{bib:op16}
  B.~A.~Dobrescu and I.~Mocioiu,
  JHEP {\bf 0611} (2006) 005
  doi:10.1088/1126-6708/2006/11/005
  [hep-ph/0605342].

\bibitem{Agnes:2018fwg} 
  P.~Agnes {\it et al.} [DarkSide Collaboration],
  Phys.\ Rev.\ D {\bf 98}, no. 10, 102006 (2018)
  doi:10.1103/PhysRevD.98.102006
  [arXiv:1802.07198 [astro-ph.CO]].

\bibitem{bib:guo}
  W.~L.~Guo, Z.~L.~Liang and Y.~L.~Wu,
  Nucl.\ Phys.\ B {\bf 878} (2014) 295
  doi:10.1016/j.nuclphysb.2013.11.016
  [arXiv:1305.0912 [hep-ph]].

\bibitem{Hoferichter:2016nvd} 
  M.~Hoferichter, P.~Klos, J.~Menéndez and A.~Schwenk,
  Phys.\ Rev.\ D {\bf 94}, no. 6, 063505 (2016)
  doi:10.1103/PhysRevD.94.063505
  [arXiv:1605.08043 [hep-ph]].
  


  \bibitem{Catena:2015uha} 
  R.~Catena and B.~Schwabe,
  ``Form factors for dark matter capture by the Sun in effective theories,''
  JCAP {\bf 1504}, no. 04, 042 (2015)
  doi:10.1088/1475-7516/2015/04/042
  [arXiv:1501.03729 [hep-ph]].
  
 
  

  
\bibitem{bib:ds10}
  T.~Alexander {\it et al.} [DarkSide Collaboration],
searches,''
  Astropart.\ Phys.\  {\bf 49} (2013) 44
  doi:10.1016/j.astropartphys.2013.08.004
  [arXiv:1204.6218 [astro-ph.IM]].

\bibitem{bib:veto}
  P.~Agnes {\it et al.} [DarkSide Collaboration],
  JINST {\bf 11} (2016) no.03,  P03016
  doi:10.1088/1748-0221/11/03/P03016
  [arXiv:1512.07896 [physics.ins-det]].
  
\bibitem{bib:ds50_70d} 
  P.~Agnes {\it et al.} [DarkSide Collaboration],
  Phys.\ Rev.\ D {\bf 93}, no. 8, 081101 (2016)
  Addendum: [Phys.\ Rev.\ D {\bf 95}, no. 6, 069901 (2017)]
  doi:10.1103/PhysRevD.93.081101, 10.1103/PhysRevD.95.069901
  [arXiv:1510.00702 [astro-ph.CO]].

\bibitem{Response} 
  P.~Agnes {\it et al.} [DarkSide Collaboration],
  JINST {\bf 12}, no. 12, T12004 (2017)
  doi:10.1088/1748-0221/12/12/T12004
  [arXiv:1611.02750 [physics.ins-det]].

\bibitem{SCENE} 
  H.~Cao {\it et al.} [SCENE Collaboration],
  Phys.\ Rev.\ D {\bf 91}, 092007 (2015)
  doi:10.1103/PhysRevD.91.092007
  [arXiv:1406.4825 [physics.ins-det]].
  
\bibitem{ARIS} 
  P.~Agnes {\it et al.},
  Phys.\ Rev.\ D {\bf 97}, no. 11, 112005 (2018)
  doi:10.1103/PhysRevD.97.112005
  [arXiv:1801.06653 [physics.ins-det]].
  
\bibitem{Agnes:2018ves} 
  P.~Agnes {\it et al.} [DarkSide Collaboration],
  Phys.\ Rev.\ Lett.\  {\bf 121}, no. 8, 081307 (2018)
  doi:10.1103/PhysRevLett.121.081307
  [arXiv:1802.06994 [astro-ph.HE]].


\bibitem{bib:ds50_50d}
  P.~Agnes {\it et al.} [DarkSide Collaboration],
Nazionali del Gran Sasso,''
  Phys.\ Lett.\ B {\bf 743} (2015) 456
  doi:10.1016/j.physletb.2015.03.012
  [arXiv:1410.0653 [astro-ph.CO]].

\bibitem{Agnes:2015qyz} 
  P.~Agnes {\it et al.} [DarkSide Collaboration],
  JINST {\bf 11}, no. 03, P03016 (2016)
  doi:10.1088/1748-0221/11/03/P03016
  [arXiv:1512.07896 [physics.ins-det]].
  


  
\bibitem{Agnes:2016nis} 
  P.~Agnes {\it et al.} [DarkSide Collaboration],
  JINST {\bf 11}, no. 12, P12007 (2016)
  doi:10.1088/1748-0221/11/12/P12007
  [arXiv:1606.03316 [physics.ins-det]].
  
\bibitem{Agnes:2017dfc} 
  P.~Agnes {\it et al.} [DarkSide Collaboration],
  JINST {\bf 12}, no. 12, P12011 (2017)
  doi:10.1088/1748-0221/12/12/P12011
  [arXiv:1707.09889 [physics.ins-det]].
  
\bibitem{Agnes:2017grb} 
  P.~Agnes {\it et al.} [DarkSide Collaboration],
  JINST {\bf 12}, no. 10, P10015 (2017)
  doi:10.1088/1748-0221/12/10/P10015
  [arXiv:1707.05630 [physics.ins-det]].
  
\bibitem{bib:uar1} D. Acosta-Kane et al., Nucl. Inst. Meth. A {\bf 587}, 46 
(2008); H. O. Back et al., arXiv:1204.6024v2 (2012); H. O. Back et al., 
arXiv:1204.6061v2 (2012); J. Xu et al., Astropart. Phys. {\bf 66}, 53 (2015).



\end{thebibliography}
\end{document}